\documentclass[twocolumn]{aastex631}

\graphicspath{{./}{figures/}{figures/photos/}}

\usepackage{verbatim}
\usepackage{animate}
\usepackage{threeparttable}

\begin{document}

\title{A Monte Carlo Simulation of the Broad Band X-ray Emission of the Accreting Millisecond X-ray pulsar MAXI J1816--195}
\correspondingauthor{Yuan You, Shuang-Nan Zhang}
\email{youyuan@ihep.ac.cn, zhangsn@ihep.ac.cn}

\author[0000-0002-0352-8148]{Yuan You}
\affiliation{Key Laboratory of Particle Astrophysics, Institute of High Energy Physics, Chinese Academy of Sciences, 19B Yuquan Road, Beijing 100049, China}

\author[0000-0001-5586-1017]{Shuang-Nan Zhang}
\affiliation{Key Laboratory of Particle Astrophysics, Institute of High Energy Physics, Chinese Academy of Sciences, 19B Yuquan Road, Beijing 100049, China}

\author[0000-0003-2310-8105]{Zhaosheng Li}
\affiliation{Key Laboratory of Stars and Interstellar Medium, Xiangtan University, Xiangtan 411105, Hunan, China}

\author[0000-0002-3776-4536]{Mingyu Ge}
\affiliation{Key Laboratory of Particle Astrophysics, Institute of High Energy Physics, Chinese Academy of Sciences, 19B Yuquan Road, Beijing 100049, China}



\begin{abstract}

MAXI J1816--195 is an accreting millisecond X-ray pulsar (AMXP) discovered in 2022. According to the Insight-HXMT data, the pulsations of this source extend all the way to over 100 keV, and its pulse profiles change from a single peak in low-energy range to double-peak in high-energy range. In this work, we simulate its energy spectra and pulse profiles with a Compton scattering Monte Carlo program. The simulation results suggest that the low energy X-ray source on the neutron star surface should be pencil-beamed radiations from the magnetic poles, and there should be a boundary layer in a hollow cylinder shape between the accretion disc and the neutron star surface: the up-scattering of the polar radiations in the boundary layer leads to the double-peak structure of the high-energy pulse profile. Under this boundary layer geometry, we suggest that the rarity of AMXPs can be caused by the smearing of the pulsed emission in the boundary layer. To estimate the mass $M$ and radius $R$ of accretion-powered millisecond pulsars whose surface radiations are badly polluted by the accretion disk and boundary layer, the impact of Compton scattering in the boundary layer on the radiation should be removed before employing the X-ray pulse profile modeling method.

\end{abstract}



\section{Introduction} \label{sec:intro}

Accreting millisecond X-ray pulsars (AMXPs) are a subclass of low-mass X-ray binaries (LMXBs), in which a millisecond pulsar (MSP) accretes matter from its binary companion, generating coherent X-ray pulsations with the same period as the pulsar rotation (see \citealp{AMXP_review1,AMXP_review2}, for reviews). All the known AMXPs are X-ray transients: their X-ray pulsations are visible only during their X-ray outbursts. Generally, the neutron star (NS) in an AMXP has relatively weak magnetic fields ranging from $10^8$ to $10^9$ Gauss, and its X-ray luminosity during an outburst remains below $\sim$10\% the Eddington luminosity. According to the recycling scenario \citep{recycling_review1,recycling_review2}, AMXPs are believed to be the progenitors of radio millisecond pulsars; in fact, some sources swing between accretion- and rotation-powered states, i.e. transitional millisecond pulsars \citep{TMP_review}. So far more than 20 AMXPs are discovered. Since that AMXPs share many similar spectra and QPO properties with LMXBs and that many NSs in LMXBs indeed rotate at millisecond periods \citep{Wijnands_1999}, a question is raised naturally that why only a small fraction of LMXBs exhibit pulsations, which has been studied extensively \citep{AMXP_review1,2018ASSL..457..149C}.

Another important aspect on studying millisecond X-ray pulsars is the measurement of their stellar mass $M$ and equatorial radius $R$. The radiation regions of millisecond X-ray pulsars are on or near the surface of the NSs. Thus, the rapid rotation of the NS surface brings about a strong Doppler effect, and the strong gravitational field of the NS produces a strong general relativity effect, both of which have significant impacts on the radiations emitted from the NS surface. Therefore, the mass and radius information are ``encoded" in the pulse profile \citep{ppmodeling_o}. With estimations of $M$ and $R$, the formation and evolution of NSs can be studied and the equation of state of cold dense matter can be constrained \citep{2016ApJ...820...28O, 2016EPJA...52...18S, 2019ApJ...887...48B}. At present, a mature method of pulse profile modeling has been developed to constrain the mass $M$ and radius $R$ of a millisecond pulsar \citep{ppmodeling_oblate, ppmodeling_review, ppmodeling_nicer2}. This method can simultaneously estimate both $M$ and $R$ of rotation-powered millisecond pulsars with high accuracy \citep{ppmodeling_nicer3, ppmodeling_rpmp_e1}, but is not as accurate when employed on accretion-powered millisecond pulsars because of the increase of parameter number and the uncertainty of polar cap model \citep{ppmodeling_amxp_e3, ppmodeling_amxp}.

MAXI J1816--195 was discovered by MAXI/GSC on 2022 June 7 as an X-ray transient \citep{1816ATel}. In the following observations, NICER discovered its 528 Hz X-ray pulsations and thermonuclear bursts, confirming that MAXI J1816--195 is an AMXP \citep{1816Bult2022}. According to the observations of Insight-HXMT during the 2022 burst, MAXI J1816--195 showed obvious pulsed emission from 2 keV up to $>95$ keV \citep{1816Li2023}. This was not the first time such high energy pulsed emission in AMXPs has been detected: the detection of pulsed emission above $>45$ keV was reported with Insight-HXMT data in observations of Swift J1756.9--2508 in its 2018 burst \citep{1756Li2021}; the detection of pulsations up to $>90$ keV was also reported with INTEGRAL data in observations of IGR J00291+5934 \citep{AMXP_INTEGRAL}. Because of the weak NS surface magnetic fields of AMXPs, it is believed that the intrinsic radiation of AMXPs cannot reach $\sim100$ keV, therefore, there must be some special photon hardening mechanisms present. Studying this radiation mechanism will help us to reveal the accretion structure and radiation mechanism of AMXPs, so that the intrinsic pulse profiles of the central MSPs can be better modelled and eventually the mass $M$ and radius $R$ of these NSs can be constrained more accurately.

In this work, we present the X-ray Monte Carlo simulation of the broad band X-ray emission of MAXI J1816--195 to reproduce the observed main features. We summarize the timing and spectral X-ray radiation characteristics of this source and make an AMXP geometrical model containing a boundary layer in Section~\ref{sec:obs}. In Section~\ref{sec:MC}, we describe the Compton Scattering Monte Carlo simulation method we used, then we show the parameter values of the geometrical model obtained from the simulation and the accuracy of the simulation compared with X-ray observation data. In Section~\ref{sec:discussion} we discuss, based on the boundary layer geometrical model established by the simulation,  the possible reasons of the rarity of AMXPs, and the influence of Compton scattering on the estimations of NS mass $M$ and radius $R$.

\section{X-ray observation characteristics and the geometrical model} \label{sec:obs}

In this work, for the X-ray radiation characteristics of MAXI J1816--195, we mainly use the data and results reported in \citealt{1816Li2023}, in which broadband timing and spectral observations of NICER, NuSTAR and Insight-HXMT are summarized. Based on these results, we establish a geometrical model of the accretion-powered millisecond pulsar, which will be tested by simulation in Section~\ref{sec:MC}.

\subsection{Pulse features}

\begin{figure*}
\centering
\includegraphics[width=18cm]{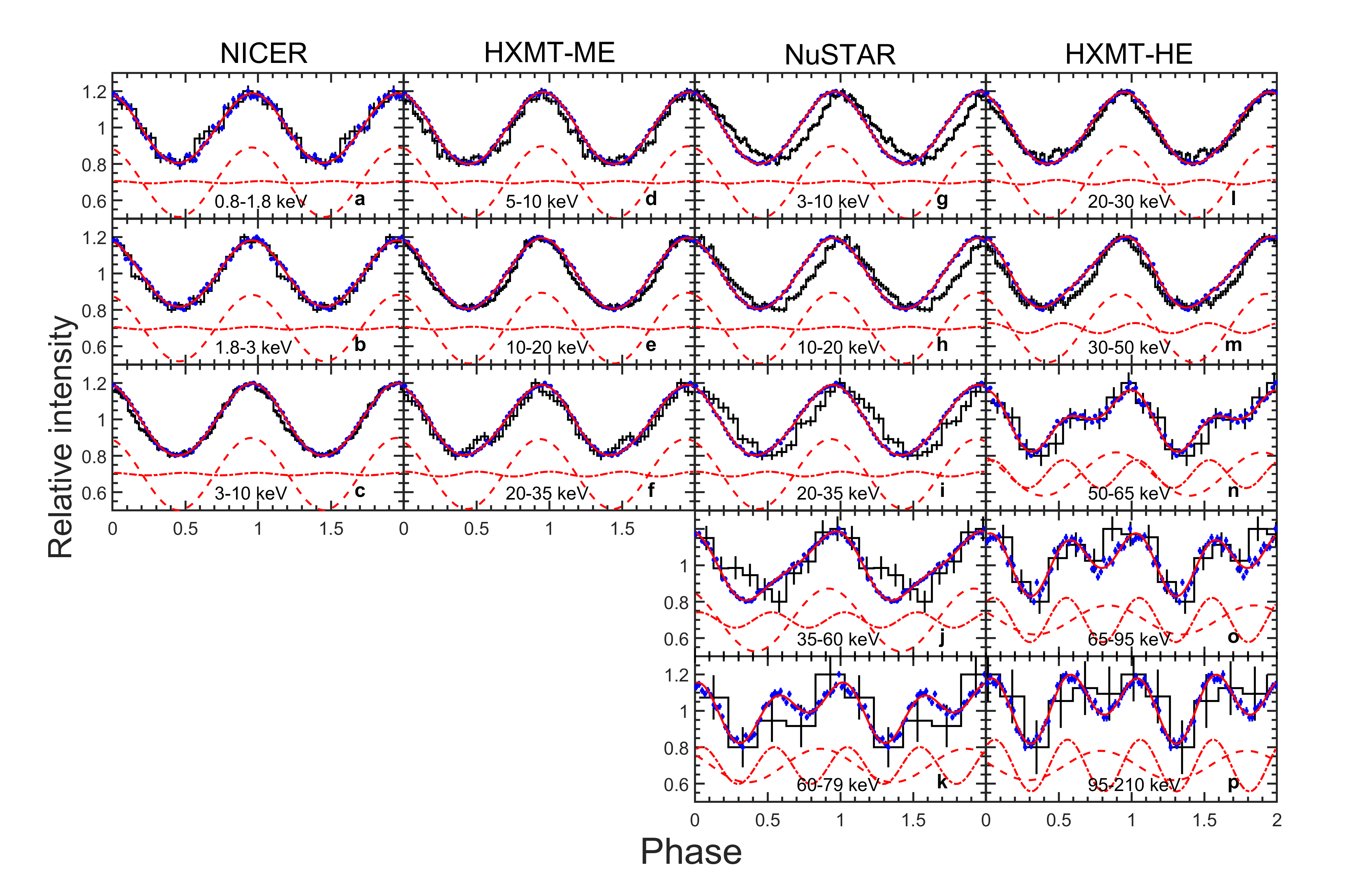}
\caption{The pulse profiles of MAXI J1816--195 in different energy bands and observed with different instruments. The black histograms are the observed results of NICER (panels a$-$c, $0.8-10$ keV), NuSTAR (panels g$-$k, $3-79$ keV) and Insight-HXMT (panels d$-$f, $5-35$ keV for ME; panels l$-$p, $20-210$ keV for HE), respectively. The blue lines are the corresponding Fourier fitting pulse profiles according to Equation~\ref{eq:Fourier}, with cyan lines the fundamental overtones and green lines the first overtones. The red lines are the pulse profiles generated by MC simulation (see Section~\ref{sec:MC} for description). The observed profiles and simulated profiles are normalized between 0.8 and 1.2. The amplitudes of fundamental and first overtones scale proportionally with the total profiles, while their average intensities are shifted to 0.7 for clarity. In order to improve clarity, two cycles are shown. The error bars represent $1\sigma$ errors. The observed pulse profiles are taken directly from \citealp{1816Li2023}.}
\label{fig:waveforms}
\end{figure*}

The pulse profiles of MAXI J1816--195 during its 2022 outburst in different energy bands observed by NICER, NuSTAR and Insight-HXMT are shown in Figure~\ref{fig:waveforms}. The pulse profiles under 50 keV exhibit an single peaked shape; however, a second but weaker pulse seems to pop-up at energies above $\sim50$ keV. To quantitatively confirm the morphology changes of the pulse profiles as a function of energy, a series of Fourier decomposition of these pulse profiles are made by,
\begin{equation}
F(\phi) = A_0 + \sum_{k=1}^2 A_k \ \cos[2\pi\ k (\phi-\phi_k)],
\label{eq:Fourier}
\end{equation}
where $A_{1}$ and $A_{2}$ are the amplitudes, $\phi_1$ and $\phi_2$ are the phase angles (in radians/$2\pi$), of the fundamental and the first overtone, respectively, and $A_0$ is the constant level of the profile. These fundamental pulses and first overtones are also plotted in Figure~\ref{fig:waveforms}. The Fourier decomposition results show that the pulsed fraction increases with energy in soft X-ray energy range until reaching its peak in $\sim20-30$ keV. The amplitudes of the fundamental has a significant decrease above $\sim35$ keV, while the relative amplitude of the first overtone compared with the fundamental overtone increases with energy, making the total pulse profile change to the double-peak shape.

\begin{figure}
\centering
\includegraphics[width=9cm]{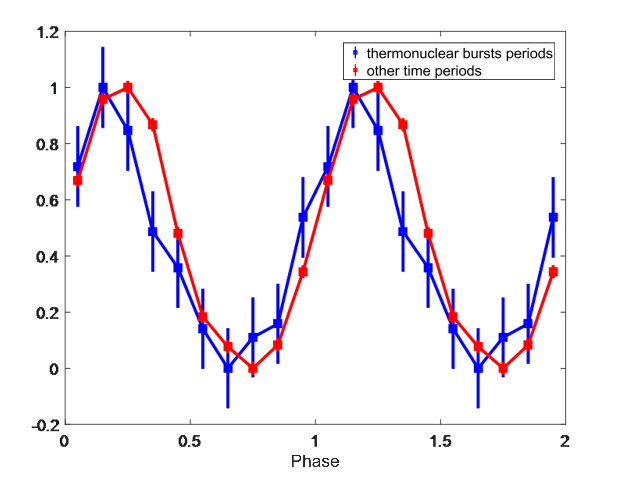}
\caption{The pulse profiles of MAXI J1816--195 in $1-10$ keV during its 2022 outburst observed by NICER. The blue line corresponds to the pulse profile during thermonuclear bursts and the red line corresponds to pulse profile when there is no thermonuclear burst. In order to improve clarity, two cycles are shown. There is no statistically significant deviation between the two pulse profiles.}
\label{fig:NICER_tburst}
\end{figure}

It is known that the pulsation of a pulsar should originate from the low energy X-ray source on the surface of the NS, usually to be one or several hot spots. If the low energy X-ray source is a hot spot, then the pulse profile should be effected by thermonuclear bursts unless the hot spot is very special. This is because during a burst the emission region of thermal radiation will extend to a large part of NS surface that can very likely include the hot spot area. MAXI J1816--195 is a prolific thermonuclear burst source. With NICER data, we check its pulse profile during its thermonuclear burst, as shown in Figure~\ref{fig:NICER_tburst}. There is no significant change in the pulse profiles of $<10$ keV energy range during and out of thermonuclear bursts, which means that the hot spot on MAXI J1816-195 is not influenced by the bursts and thus quite special. We suggest that this special ``hot spot" is actually the magnetic pole of the NS: the accreted material accumulates at the magnetic pole to form mounds, which do not burn out during thermonuclear bursts, so that thermonuclear bursts have no significant impact on the X-ray pulse profile. In this scenario, similar to other magnetised accreting pulsars, it is natural to assume that the magnetic pole X-ray source is a cut-off power-law radiation source generated by the collision of accreted matter onto the NS surface, rather than a pure blackbody radiation source. In our subsequent simulations, we also find that the hot spot temperature required to generate the observed pulse profiles is too high compared to NS surface temperature, which also disfavors the normal hot spot scenario but prefers the magnetic pole scenario. We will discuss it in details in section~\ref{sec:doublepeak}.

\subsection{Spectral features}

According to \citealt{1816Li2023}, the best-fit Xspec phase-averaged spectral model of MAXI J1816--195 is {\tt TBabs(gaussian + gaussian + diskbb + compps)} with reduced $\chi^{2}$ 1.11. In the view of \citealp{1816Li2023}, the {\tt compps} component in this spectral model corresponds to a slab disk corona; however, this scattering geometry fails to explain the observed high energy pulsations. The solid angle of a slab disk corona towards a low energy X-ray source on the NS surface is too small, so that almost all photons in the {\tt compps} component are originally from the disk. The accretion disk cannot generate pulsations with the same frequency as the NS rotation, therefore, the {\tt compps} component should not have obvious pulsations. However, in the energy range of $>10$~keV where the {\tt compps} component is dominant, the peak value of the pulse fraction appears. Despite of this, this Xspec spectral model still provides two valuable pieces of information. One is that there is an accretion disk emitting blackbody radiation, the disk temperature at inner radius $\sim 0.63$ keV given by Xspec can be a reference value in our simulations. The other is that the high energy radiations are Comptonized radiations, thus the double-peaked pulsation in high energy range should be formed by a special scattering geometry. Since that the {\tt diskbb} component is dominant in soft X-ray bands, the accretion disk should not be obscured. Therefore, the possibility of a hot disk corona as the up-scattering Comptonization region of the NS surface radiation should be excluded.

\subsection{The geometrical model}

\begin{figure}
\centering
\includegraphics[width=9cm]{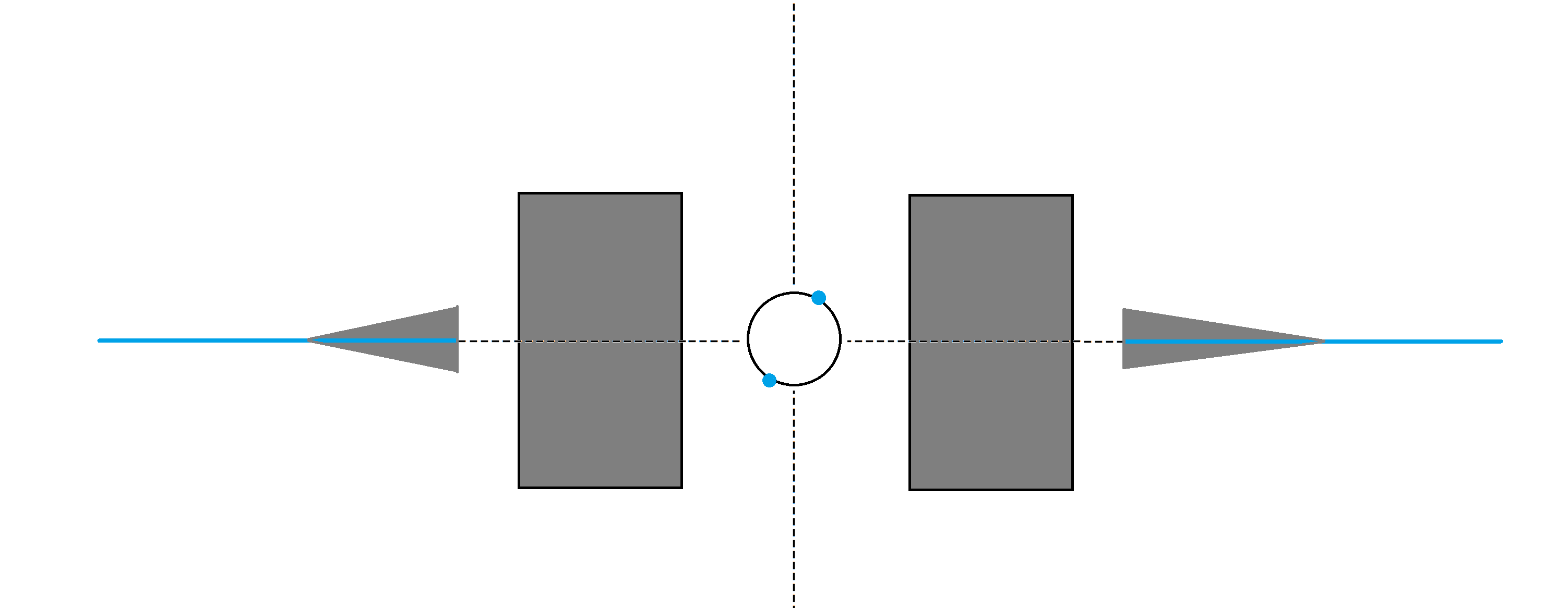}
\caption{A sketch of the geometrical model used in MC simulations. The scattering regions are marked by grey color and the low energy X-ray sources are marked by blue color. There is a hollow cylinder boundary layer between the NS and accretion disk, scattering photons emitted from the NS. The two magnetic poles of the NS are two point low energy X-ray sources and the accretion disk is a disk low energy X-ray source. The disk corona covering the disk has optical depth decreasing with radius.}
\label{fig:geo_structure}
\end{figure}

A geometrical model is established according to the observed features mentioned above, whose structure is shown in Figure~\ref{fig:geo_structure}. This model consists of a MSP, its accretion disk and some scattering coronae. The MSP at the center is set to be a perfect sphere with radius 11 km and rotation frequency 528 Hz, and it has two symmetrical magnetic poles. There are two kinds of low energy X-ray sources in this geometrical model: the two polar low energy X-ray sources on the NS surface, and the disk low energy X-ray source on the thin accretion disk. After emitted, some of seed photons will be scattered in the coronae according to their emitted angle. We assume that in the coronae the only relevant physical process is Compton scattering between incident photons and hot electrons in the coronae; the number of photons remains conserved during scattering. The NS surface and the accretion disk are set to be the absolute absorption surface of photons: any photon that reaches these surfaces will be absorbed, i.e., we neglect any possible reflections and radiative transfer in our simulation. For photons not absorbed, they will fly out of the system and finally be detected by distant observers at their respective observing inclination angles. Due to the single-peaked shape of the pulse profile in low energy range, both the observing inclination angle and the magnetic axis inclination angle of this system should not be large; in this way, the radiation of only one magnetic pole can be observed directly, and the other pole is obscured by the NS itself, so that the pulsation is single-peaked.

Between the disk and the NS, there is a high temperature corona of hollow cylinder shape; the low energy radiation from the NS surface will be scattered by this corona and finally become the double-peaked pulsating high energy radiation. This configuration of corona can correspond to the boundary layer of the NS, which is correlated with kHz QPOs \citep{boundary_layer_kHzQPO}. The reason for not choosing a lamppost corona often used for accreting black hole X-ray binaries is that it is quite difficult for a lamppost corona to convert single-peaked radiation to double-peaked radiation by Compton scattering. In order to fit the shape of the energy spectrum, it is still possible to have a disk corona, but this disk corona should not be too thick and the temperature should not be too high, so that the disk radiation can still penetrate from it and be observed in the low energy range. To simplify the simulation, the boundary layer and the disk corona are homogeneous and do not have bulk motion, but they have different densities and temperatures.

The spectrum of two polar low energy X-ray sources on the NS surface is set to be {\tt cutoffpl} in XSPEC. Considering that AMXPs have relatively weak magnetic fields ranging from $10^8$ to $10^9$ Gauss which are not strong enough to form accretion columns, the polar low energy X-ray sources should be on the NS surface with negligible height and its radiation is thus pencil-beamed. Specifically, we simplify the polar low energy X-ray sources as two point sources at the north and the south poles of the NS, with angular distribution of seed photons as follows:
\begin{equation}
  P_{\rm{seed}} \propto \cos\theta,
\end{equation}
where $\theta\in[0,\pi/2]$ is the angle between the normal direction of the NS surface and the momentum of the seed photons; $P_{\rm{seed}}$ is the emitting probability of a seed photon. The accretion disk is set to be a thin disk, and the disk low energy X-ray source is described as {\tt diskbb} in XSPEC.

The AMXP geometrical model has been fully established by the description above; we will use this model to simulate the X-ray radiation of MAXI J1816--195, and the parameters of the geometrical model are determined by comparing the observed results with the simulation results.

\section{Monte Carlo simulation} \label{sec:MC}

\subsection{Simulation methods}

In the simulation of MAXI J1816--195, we use a Monte Carlo (MC) program we developed previously \citep{myComptonMC}. This is a Geant4 program specialized in simulating the Compton scattering between photons and free electrons: it tracks the scattering process of a photon in a flat space and accurately calculate the outgoing energy and direction of the photon. The program can construct homogeneous electron coronae of arbitrary shape, temperature and bulk velocity, making it suitable for various scattering situations.

In this simulation, for the two polar low energy X-ray sources, we first generate the pencil-beamed {\tt cutoffpl} photons in the inertial stationary frame of the NS pole, then obtain the energy and direction of photons in the reference frame of the observer by Lorentz transformation according to the velocity of the NS magnetic pole, and these photons will serve as the seed photons of the scattering process; for the disk low energy X-ray sources, the Doppler effect is ignored. Since that the polar low energy X-ray sources and the disk low energy X-ray source are independent from each other, we decide to simulate these two kinds of low energy X-ray sources and their scattering process separately. After the Monte Carlo simulation of the two kinds of seed photons, we add them together with an appropriate proportion to form the spectral and timing results. The parameter values of the model are carefully selected so that the main observed features, especially the single-peaked profile at low energy and the double-peaked profile at high energy, are reproduced. However, some details are not matched between the observations and MC simulations yet. Since we do not fit the simulated results to the observed data, we cannot estimate the parameter errors and also cannot assure that the selected parameter set is the only one that can reproduce the observed features. Therefore the work presented here serves as a proof of principle of our geometrical model.


In order to get sufficient statistics, the number of resulted total photons with energy $>40$ keV is no less than $10^9$ in the simulation for the polar low energy X-ray source. Similarly, and the number of resulted total photons with energy $>20$ keV is also no less than $10^9$ for disk low energy X-ray source simulation. When counting the simulation results, the range of the outgoing directions of photons is chosen as $\pm3^{\circ}$ around the observing inclination angle. Utilizing the central symmetry of the geometrical model, radiation in the reversed direction is flipped by 180 degrees to double the statistics. It is obvious that for pulse observation, a rotating low energy X-ray source with a stationary observer is equivalent to a stationary low energy X-ray source with an orbiting observer. Therefore, the polar low energy X-ray sources on the NS are simulated in only one rotating phase. After the simulation, when counting photons to generate pulse profiles, the corresponding pulse phase of each photon is calculated by the ultimate azimuth angle and the scattering time of the photon. We stipulate that the observer observes the NS system from north to south, the NS rotates counterclockwise in the observer's view, and the 0 pulse phase is set at the moment when the azimuth angle of the northern magnetic pole is equal to the observer's azimuth angle.

\subsection{Simulation results in general}

As described above, we make assumptions about the geometry of MAXI J1816--195 that there is a boundary layer of hollow cylinder shape between the spherical NS and the multi-black body accretion disk, on two magnetic poles of the NS there are two pencil beamed low energy X-ray sources with spectra {\tt cutoffpl}, and the disk coronae has optical depth decreasing with its radius. A sketch of the geometrical model is shown in Figure~\ref{fig:geo_structure}. After careful adjustments and testing, we have obtained a set of model parameters that can reproduce the main features of the observed data, which are shown in Table~\ref{tab:model_parameters}. The detailed analyses of timing and spectral data are made in the following paragraphs.

\begin{table}[h]
  \centering
  \caption{Parameters in the MC simulation model of MAXI J1816--195.}
  \label{tab:model_parameters}
  \begin{threeparttable}
  \begin{tabular}{ccc}
    \hline
    Class & Parameter & Value \\
    \hline
    neutron star & radius & 11 km \\
                 & magnetic inclination & $30^{\circ}$ \\
                 & cut off energy & 8 keV \\
                 & photon index & -0.8 \\
                 & observing inclination & $20^{\circ}$ \\
    boundary layer & inner radius & 16 km \\
                   & outer radius & 28 km \\
                   & height & 24 km \\
                   & density & 0.8\tnote{*} \\
                   & temperature & 45 keV\\
    accretion disk & inner radius & 35 km \\
                   & temperature & 0.9 keV \\
    disk corona & inner radius & 35 km \\
                & outer radius & 835 km \\
                & temperature & 30 keV \\
                & largest height & 2 km \\
                & density & 4\tnote{*} \\
    \hline
  \end{tabular}
  \begin{tablenotes}
    \footnotesize
    \item[*] The unit is optical depth (Thomson cross section) per km.
  \end{tablenotes}
  \end{threeparttable}
\end{table}

\begin{figure}
\centering
\includegraphics[width=9cm]{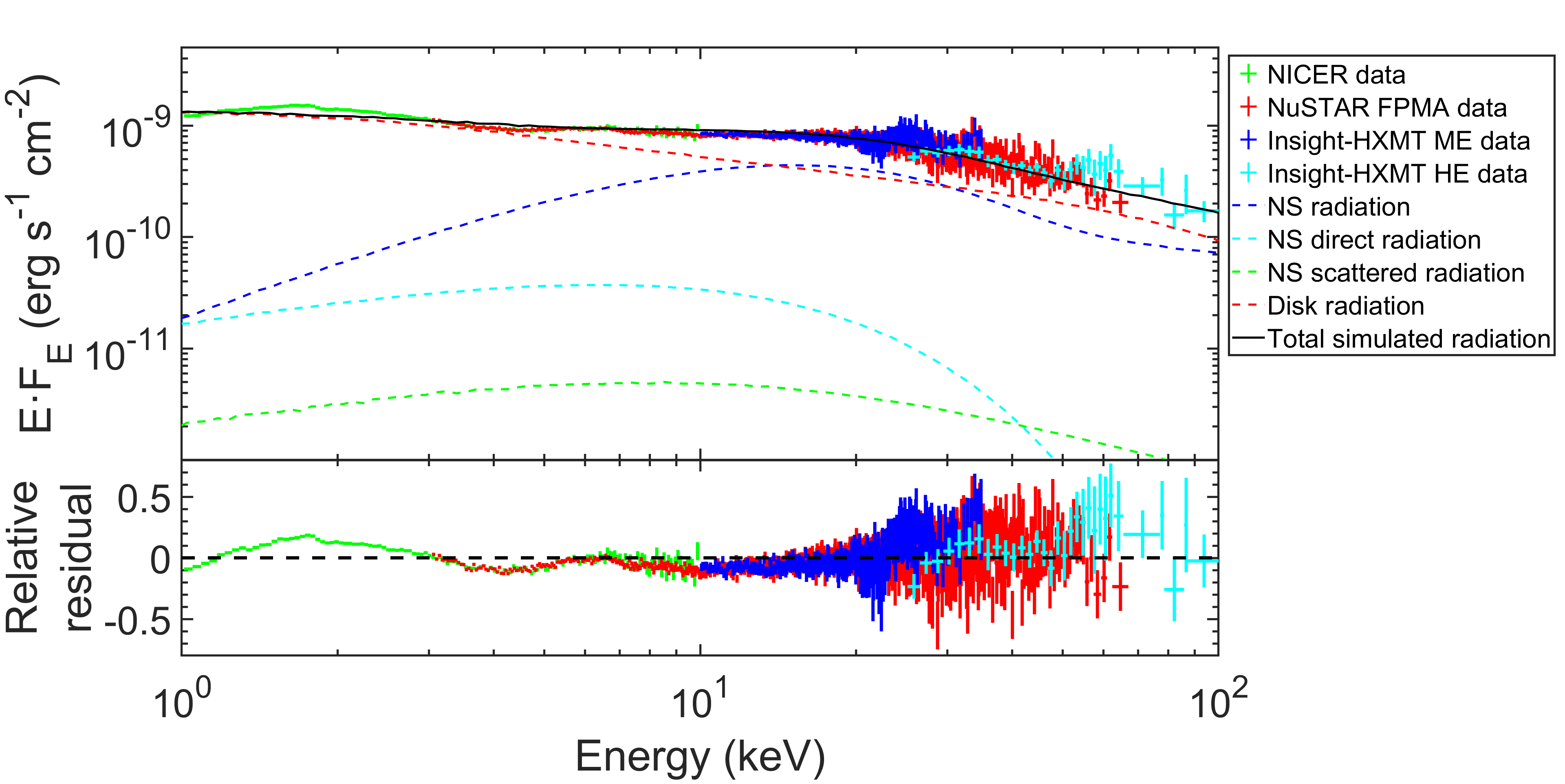}
\caption{The simulated spectra of MAXI J1816--195 compared with the observed broad band data. The observed data of NICER, NuSTAR and Insight-HXMT corresponds to crosses, the total spectrum of the simulation result is shown by the black solid line and several spectral components are shown by dashed lines. The bottom panel shows the relative residuals, i.e. the fraction between the absolute residuals and the observed data.}
\label{fig:sim_spectra}
\end{figure}

The simulation result of the phase-averaged spectrum is shown in Figure~\ref{fig:sim_spectra}. Generally, our MC simulation well reproduce the observed broadband X-ray energy spectrum of $1-100$ keV; however, because of the lack of two \textbf{gaussian} components compared with the result of \citealp{1816Li2023}, our spectrum has some minor residuals in $1-10$ keV. At energy $>50$ keV, our spectrum fits observed data better than the spectral fitting in \citealp{1816Li2023}. This is because in \citealp{1816Li2023} all scattering radiations are roughly represented by a single {\tt compps} component; however, in our model, the scattering in disk corona and the scattering in boundary layer are two components, so that the shape of the simulated spectrum is more flexible.

\begin{figure}
\centering
\includegraphics[width=9cm]{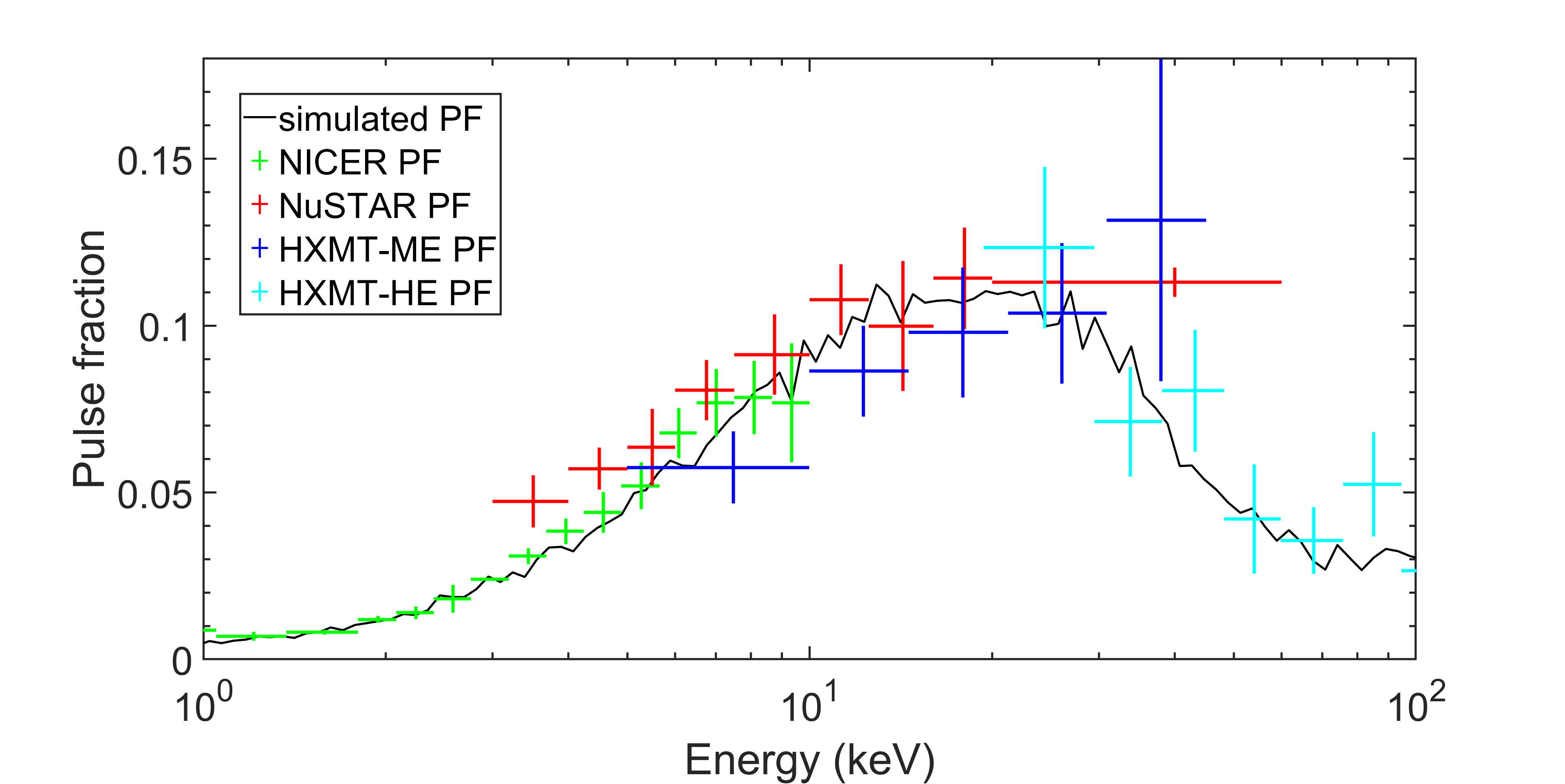}
\caption{The PF of MAXI J1816--195 as a function of energy. The observed PF of NICER, NuSTAR and Insight-HXMT corresponds to the crosses,  and the PF of simulation result is shown by the black solid line.}
\label{fig:sim_pf}
\end{figure}

In Figure~\ref{fig:sim_pf}, the variation of pulse fraction (PF) with energy is calculated, in which the PF is calculated by
\begin{equation}
  PF = \frac{f_{\rm{max}}-f_{\rm{min}}}{f_{\rm{max}}+f_{\rm{min}}},
  \label{eq:pf}
\end{equation}
where $f_{\rm{max}}$ and $f_{\rm{min}}$ are the maximum and the minimum of the X-ray flux in one pulse period, respectively. In the MC simulation, the PF value increases with energy below $10$ keV, reaches a plateau at energy $10-30$ keV and then decreases with energy above $30$ keV, which is consistent with the observations.

It is known that the observed X-ray pulsation originates from the rotation of the NS. Therefore, among the three radiation components shown in Figure~\ref{fig:sim_spectra}, the radiations directly from the NS surface has the largest PF, followed by the scattered neutron star radiations; these two components together constitute the total radiation of the NS. On the contrary, the disk radiations as the third component has no pulse at all. Due to the high proportion of the total radiation of the NS in the energy range of $10-30$ keV, a high PF plateau is formed in this energy range.

\begin{figure*}
\centering
\includegraphics[width=18cm]{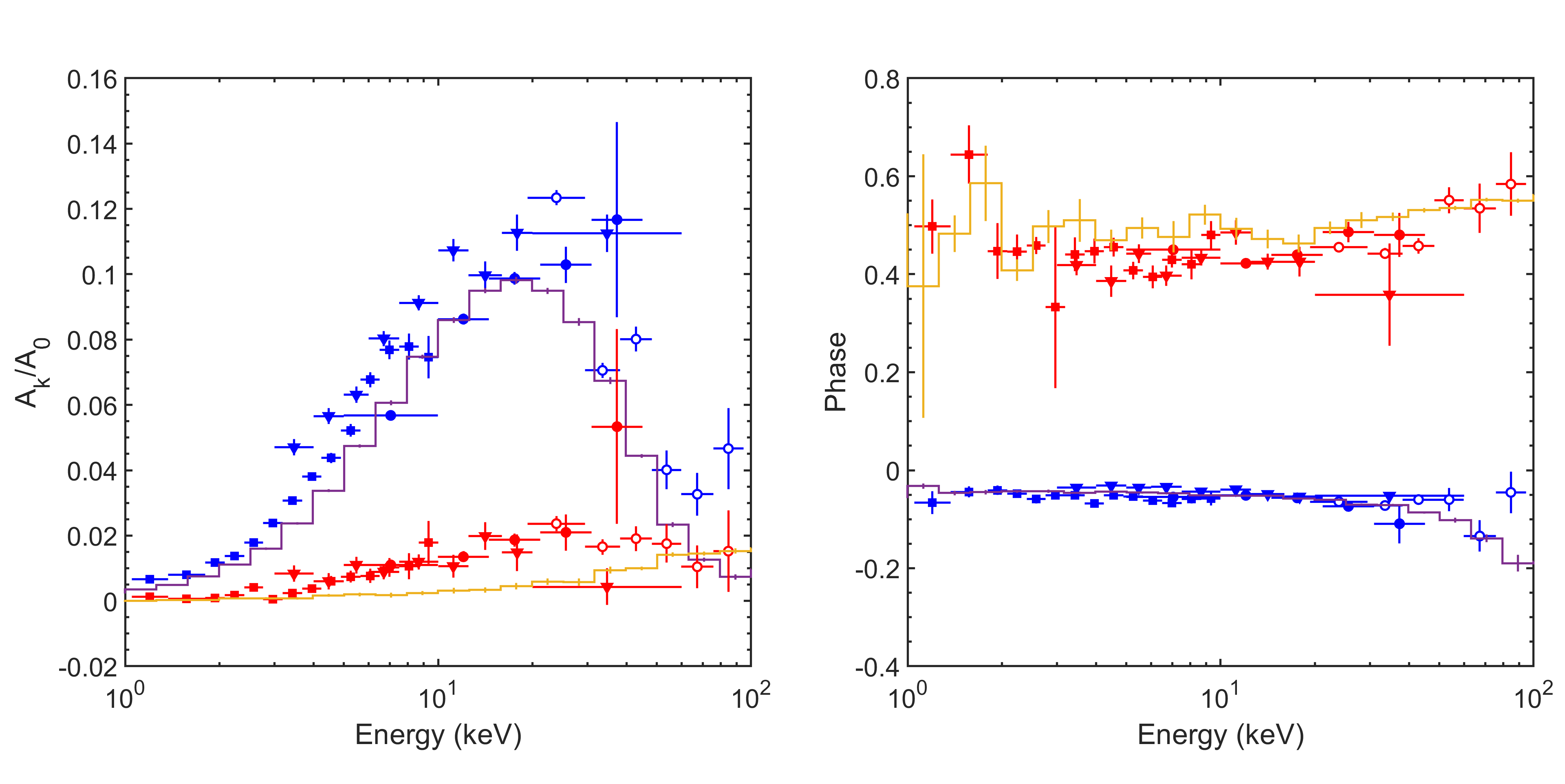}
\caption{The Fourier decomposition parameters of the observed and simulated pulse profiles of MAXI J1816--195. The decomposition equation is shown in Equation~\ref{eq:Fourier}. The left panel shows the fractional amplitude of the fundamental overtone (blue points as the observed and magenta line as the simulated) and the first overtone (red points as the observed and yellow line as the simulated) as a function of energy. The right panel shows the phase position of the fundamental overtone and first overtone as a function of energy, with the color settings the same as the left panel. The square, triangle, solid circle and circle points represent data of NICER, NuSTAR, HXMT-ME and HXMT-HE, respectively.}
\label{fig:fourier_paras}
\end{figure*}

\begin{figure}
\centering
\includegraphics[width=9cm]{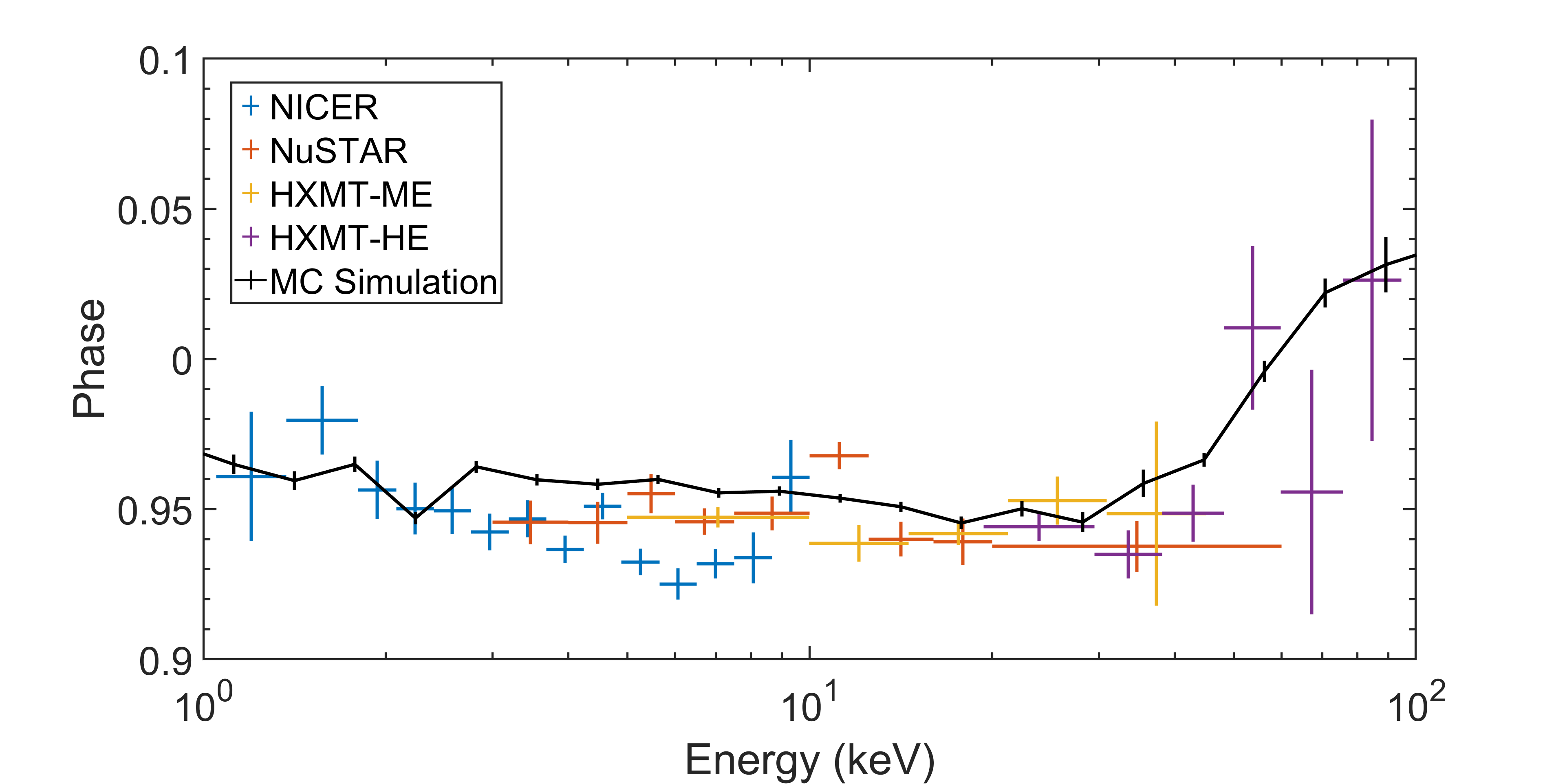}
\caption{The phase position of the highest point of the main peak in the Fourier recomposed pulse profiles (using Equation~\ref{eq:Fourier}) as a function of energy. The crosses are observation results of different instruments and the solid line is the MC simulation result.}
\label{fig:peak_phase}
\end{figure}

The simulated pulse profiles of different energy bands are also shown in Figure~\ref{fig:waveforms}. It is shown that the MC simulations reproduce well the changes of pulse shapes from low energy to high energy. Further, the Fourier decomposition results of the observed and the simulated pulse profiles according to Equation~\ref{eq:Fourier} are compared in Figure~\ref{fig:fourier_paras}, and the trend of their Fourier parameters are consistent in general. Figure~\ref{fig:peak_phase} shows the phase positions of the main peak in the pulse fraction as a function of energy. The simulation result indicates an obvious phase lag at energy $>30$ keV, which is caused by the Compton up-scattering in the boundary layer; however, the phase lag of observed data have large errors due to insufficient statistics. If confirmed by observations with high accuracy in the future, this large abd positive phase lag can serve as good evidence that the hard X-ray radiation of AMXPs are caused by the scattering in the boundary layer rather than in the accretion column or NS atmosphere as argued in some earlier models (see, for example, \citealt{ppmodeling_nicer3, ppmodeling_oblate}), because the latter cannot produce distinguishable phase lag considering the short scattering path length. Of course, this phase lag can also help constrain the geometry of the boundary layer more accurately.

\subsection{Mechanism of pulse shape changing with energy}\label{sec:doublepeak}

It is found that the transition energy from the single-peaked pulse to the double-peaked pulse is strongly related with the spectral truncation energy of the polar low energy X-ray sources on the NS surface. The simulation result of this truncation energy is about 6.5 keV, which is too high for the temperature of a normal hot spot on a NS surface. This result confirms our assumption that the low energy X-ray source on the NS surface should be the accumulated mound of accreted matters, rather than thermal radiations from a normal hot spot away from the magnetic poles. It is indicated by the spectral simulation that at low energy range, the scattered radiation is much weaker than the NS's direct radiation, thus the low energy pulse is mainly contributed by the NS's direct radiation. Due to the small inclination angle, only the northern pole can be observed directly, so that in low energy range there is only one peak which locates at phase around 0 in a pulse period. The key lies in how to understand the double-peaked pulse shape at high energy range.

\begin{figure*}
\centering
\includegraphics[width=18cm]{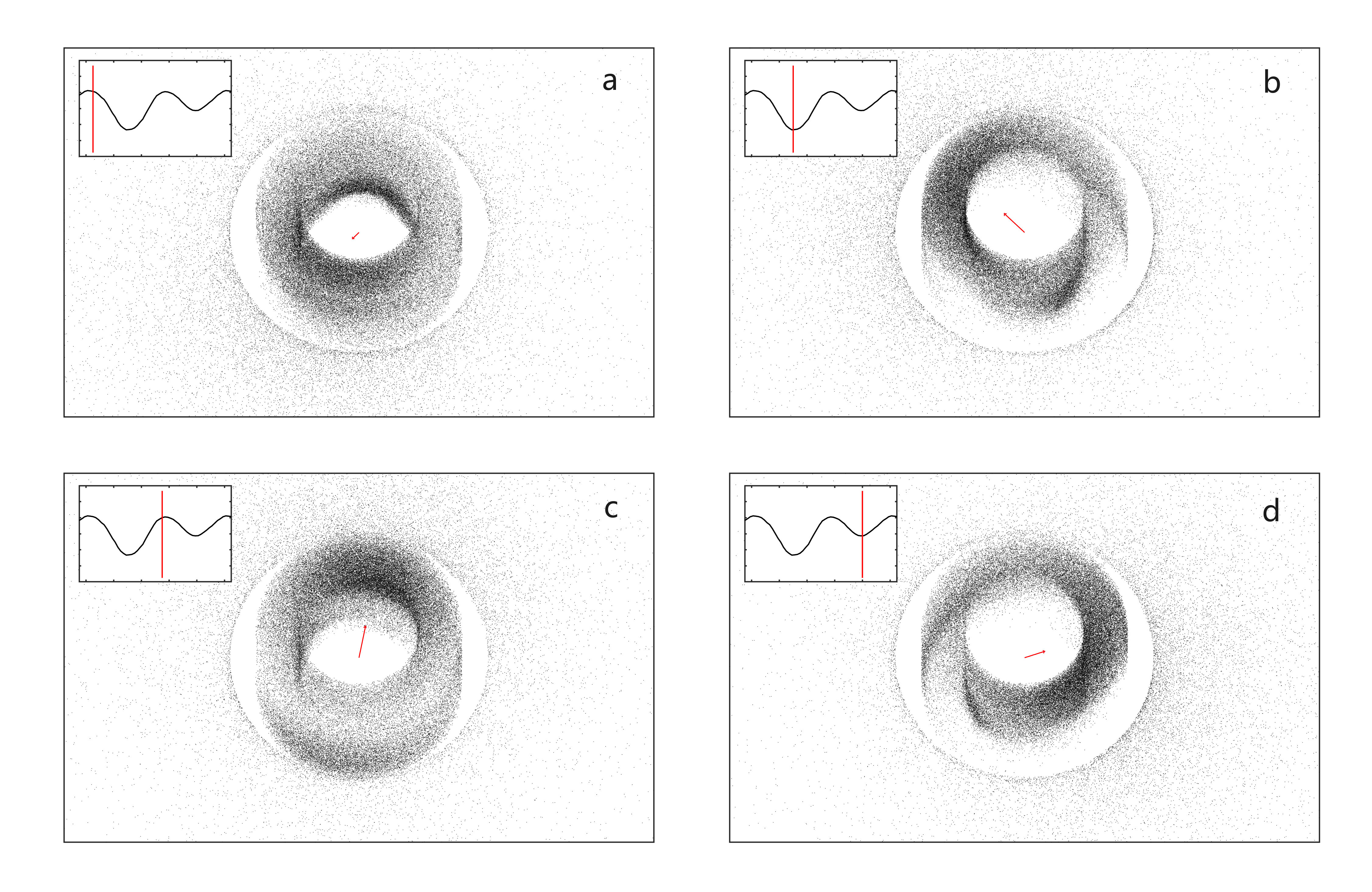}
\caption{The simulated images of MAXI J1816--195 in the energy range of $>60$ keV at different phase positions as the view of the observer on the earth (observing inclination $20^\circ$). Panels a, b, c, d corresponds to phase 0.05, 0.3, 0.55, 0.8, respectively, and the corresponding position in the pulse profile are plotted at the top left corner of each panel. In each image, every black dot corresponds to a photon, and the orientation of the northern magnetic pole of the NS is marked by the red arrow. Only photons emitted from the NS poles and scattered in the boundary layer or disk corona are drawn in these images; photons emitted from the disk and photons directly from the NS's poles without scattering are omitted.}
\label{fig:sim_photos}
\end{figure*}

\begin{figure}
\centering
\includegraphics[width=9cm]{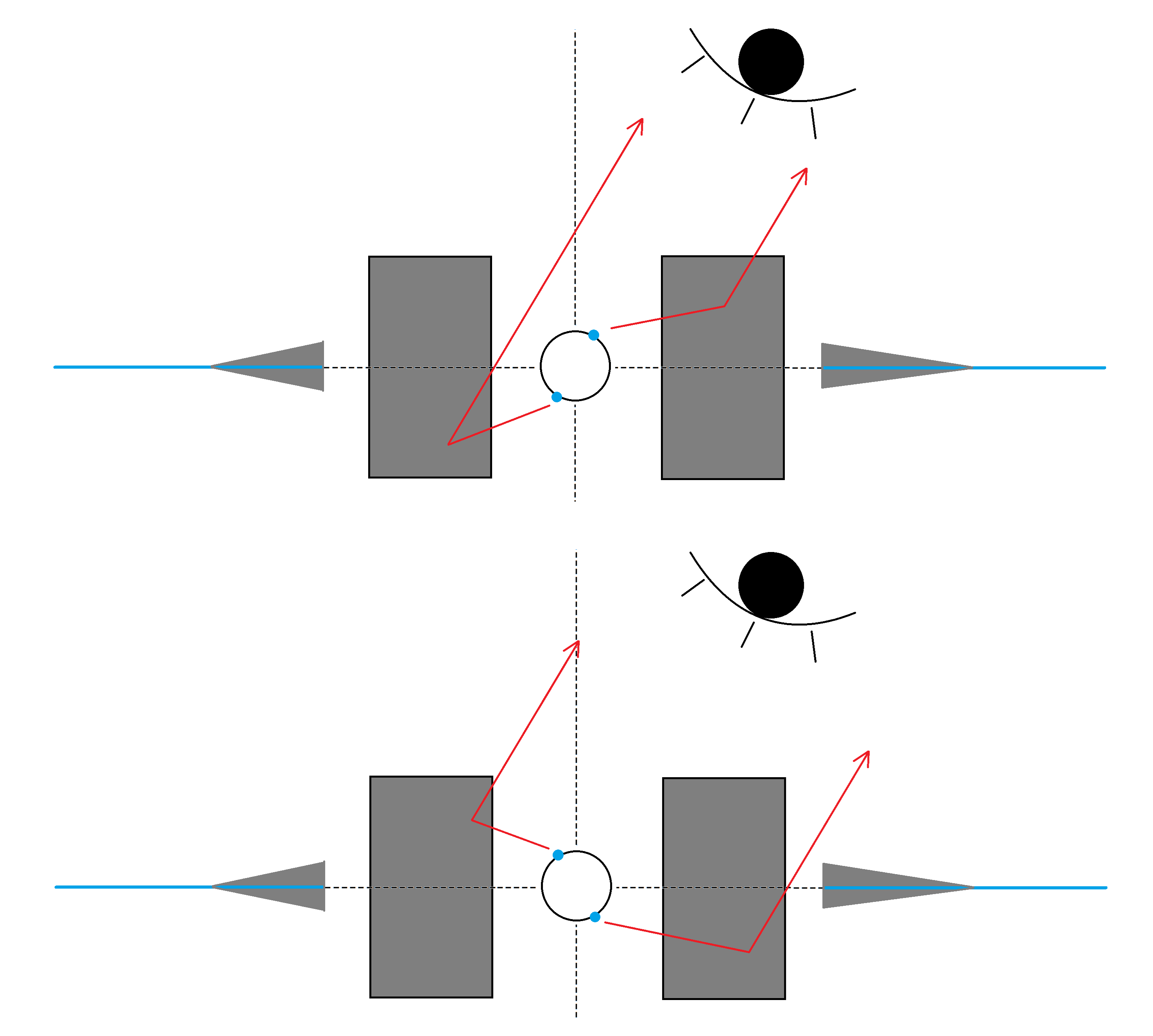}
\caption{The schematic diagrams of the photon scattering paths at different phase positions. The upper panel shows the situation of phases around 0.05, corresponds to the main peak in the pulse profile and the panel a of Figure~\ref{fig:sim_photos}; the lower panel shows the situation of phases around 0.55, corresponds to the minor peak in the pulse profile and the panel c of Figure~\ref{fig:sim_photos}.}
\label{fig:doublepeak}
\end{figure}

In order to study the generation mechanism of the double-peaked pulse at high energy, the simulated observing images of the AMXP are shown in Figure~\ref{fig:sim_photos}. In these images, only photons initially emitted from the NS surface and then scattered by the boundary layer with final energy $>60$ keV are drawn; each black dot corresponds to a photon, thus the density of points in the image is proportional to the radiation intensity. Most of the NS photons are last scattered in the boundary layer, so the outline of the hollow cylinder shape boundary layer can be clearly seen in the images; a small number of photons are last scattered at the disk corona, corresponding to the dispersed points around in the images. The images show that the two NS polar X-ray sources illuminate the boundary layer, forming two bright regions on the boundary layer that rotates with the NS. At phases around 0.05, the observer can see one bright area formed by the northern polar X-ray radiations passing through the boundary layer and another bright area formed by the southern polar radiations being reflected by the boundary layer, as the light path shown in the upper panel of Figure~\ref{fig:doublepeak}; at phase around 0.55, the observer can see one bright area formed by the northern polar radiations being scattered in the boundary layer and another bright area formed by the southern polar radiations passing through the boundary layer, as the light path shown in the lower panel of Figure~\ref{fig:doublepeak}. At phases around 0.25 or 0.75, the bright area illuminated by the southern polar is obscured by the boundary layer itself, therefore the corresponding flux at these phases are smaller.

The above analysis indicates that the double-peaked pulse shape is the geometric effects of the boundary layer scattering. These two peaks are located around phase 0 and 0.5, respectively, with no significant difference in the amplitude if only the scattering in the boundary layer is concerned. When taking into account both the NS's direct radiation and the scattered radiation, the peak around phase 0 becomes higher due to superposition, which forms the main peak. Therefore, only when the flux of the scattered radiation is significantly greater than the flux of the NS's direct radiation, will the secondary peak become apparent.

\section{Discussions} \label{sec:discussion}

\subsection{The rarity of AMXPs}

To explain the phenomenon that most LMXBs usually do not show X-ray pulsations, several mechanisms have been proposed, including the burial of the NS surface magnetic field by accretion \citep{AMXP_m_burial1, AMXP_m_burial2, AMXP_m_burial3}, the smearing of pulsations due to gravitational light bending \citep{AMXP_GR_smear1, AMXP_GR_smear2}, onset of MHD instabilities at magnetospheric boundary \citep{AMXP_MHD_instable}, the alignment of the NS magnetic and rotational axes \citep{amxp_m_align1, amxp_m_align2}, and the smearing of pulsations by an optically thick corona \citep{AMXP_corona_smear1, AMXP_corona_smear2, AMXP_corona_smear3}. Due to the lack of consideration of the effects of magnetic and gravitational fields on photon scattering in our simulation program, we are only able to conduct simulation tests on the mechanisms of the alignment of the NS magnetic and rotational axes and the smearing of pulsations by an optically thick corona.

\begin{figure}
\includegraphics[width=9cm]{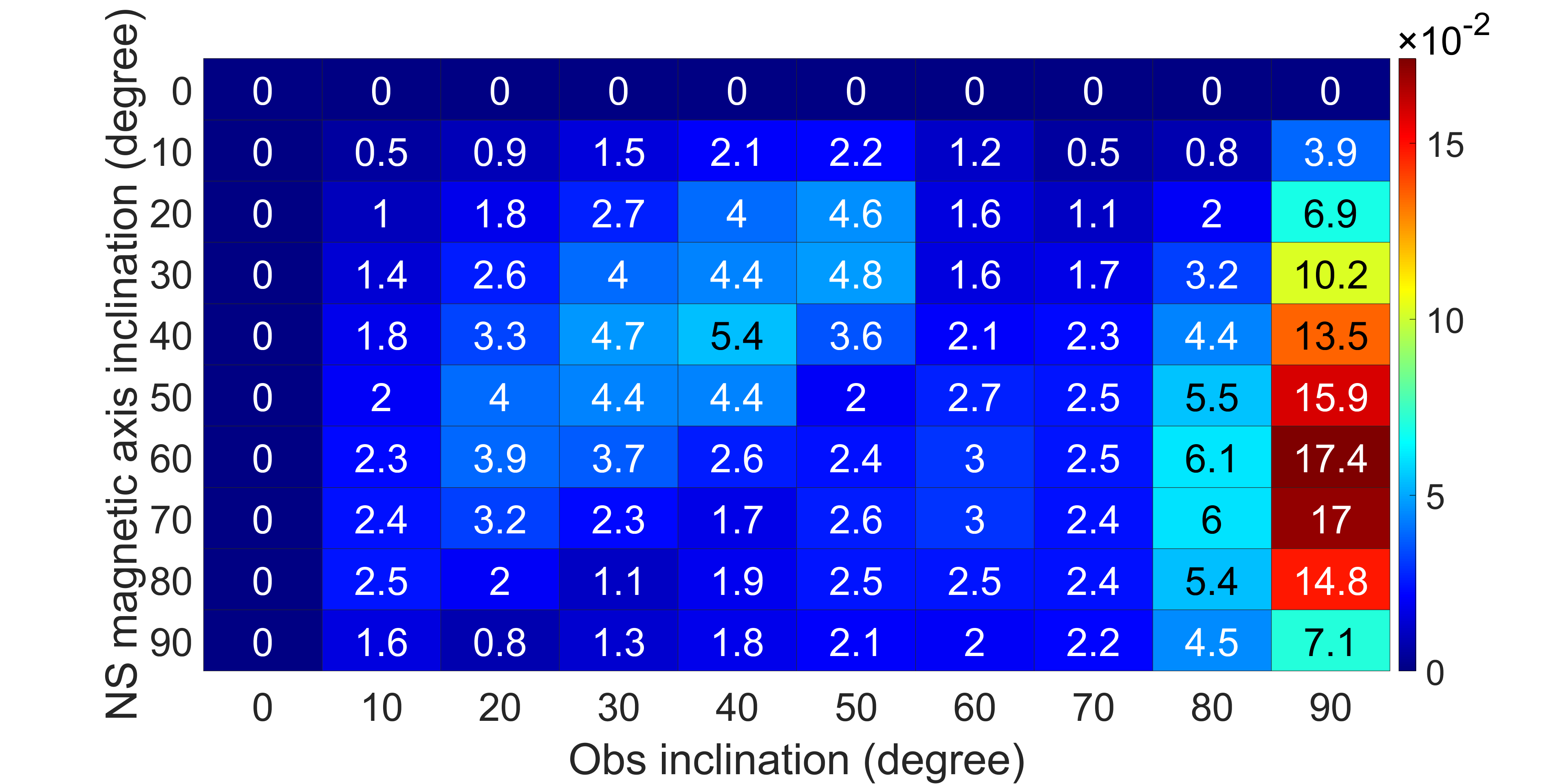}
\caption{The simulation results of pulse fraction in $1-10$ keV in cases of different magnetic inclinations and observing inclinations. The pulse fractions are calculated with Equation~\ref{eq:pf}. All these simulated cases have the same geometry parameter values with the MAXI J1816--195 simulation, except their magnetic and observing inclinations.}
\label{fig:pf_m_i}
\end{figure}

It is believed that NSs are spun-up by accretion, and during this process crustal shear stress breaks the crust of a NS into “plates” that drift towards the rotational poles \citep{ns_mag_mov1, ns_mag_mov2}. Therefore, the magnetic axes of recycled millisecond pulsars are expected to be nearly aligned with their spin axes \citep{ns_mag_mov3, ns_mag_mov4}. Based on this evolution scenario of pulsars, a nearly aligned moving spot model of AMXPs is built and tested by X-ray pulse profile modeling \citep{amxp_m_align1, amxp_m_align2}. In this model, the lack of pulsation in most LMXBs is explained by the approximately zero magnetic inclination angle, and the magnetic inclination of AMXPs are believed to be about a few degrees, so that the modelled X-ray pulse profile can coincide with the usually detected sinusoidal and low-amplitude pulse profiles. The intermittency of some AMXPs can also be explained by hot spots slightly moving around the magnetic poles, which may be caused by the turbulence of accretion rate or disk inner radius. It should be noted that in the process of building the nearly aligned moving spot model of AMXPs, the re-radiations of photons after emitted from hot spots are neglected for simplification in our simulations.

\begin{figure}
\includegraphics[width=9cm]{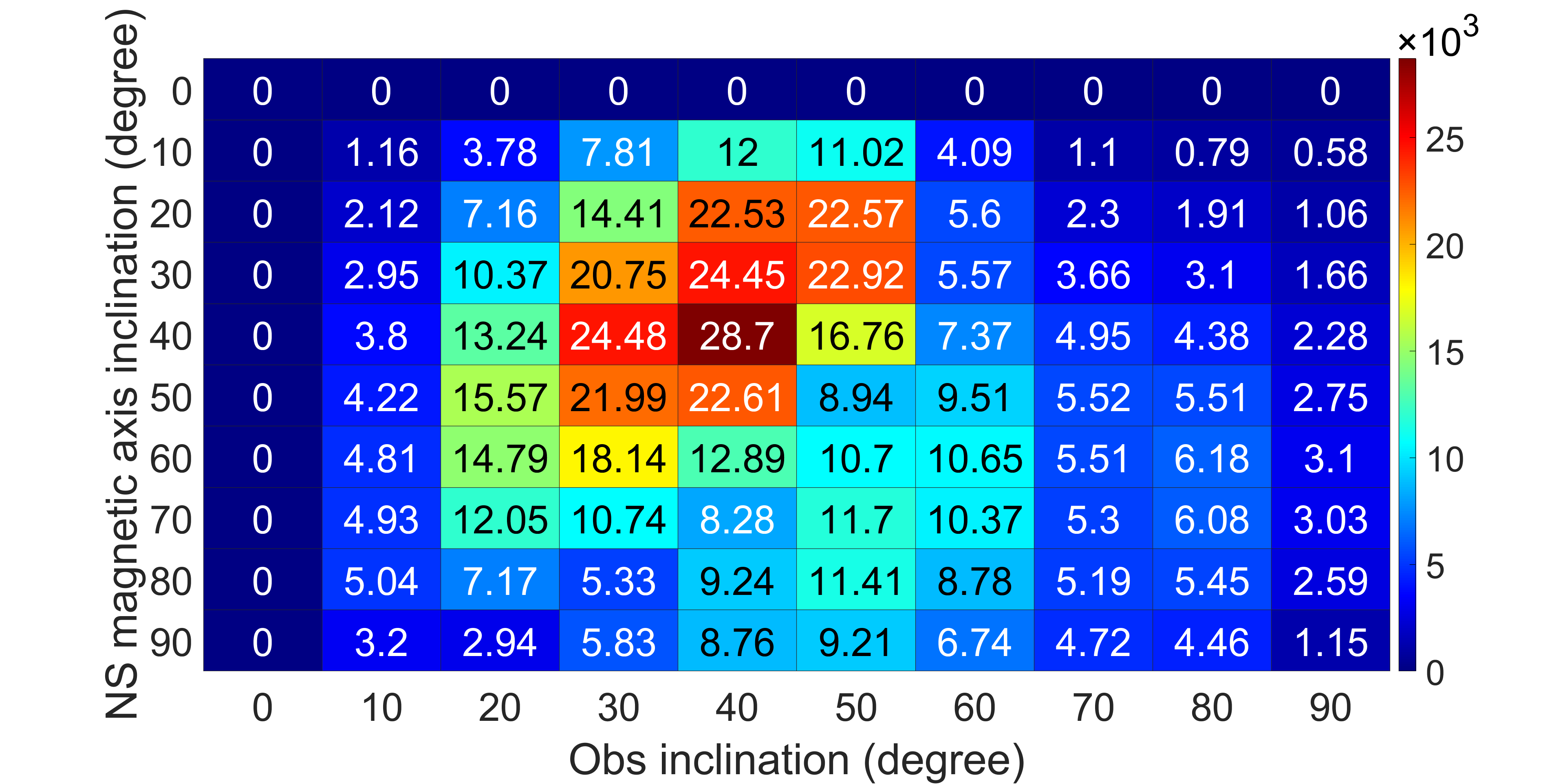}
\caption{The simulation results of pulse amplitude in $1-10$ keV in cases of different magnetic inclinations and observing inclinations, in units of simulated photon number. All these simulated cases have the same geometry parameter values with the MAXI J1816--195 simulation, except their magnetic and observing inclinations.}
\label{fig:amplitude_m_i}
\end{figure}

If the nearly aligned moving spot model of AMXPs is correct, all radio MSPs as the evolutionary products of AMXPs should have very small magnetic inclinations. However, the observation results of radio pulsars shows a wide distribution of magnetic inclinations \citep{radioMSP}. Obviously, MAXI J1816--195 does not match with the AMXP model of nearly aligned moving spot, because the pulse fraction ($>10\%$ at energy $10-30$ keV) are larger than the expectation of this AMXP model. More importantly, the pulse fraction of the pulsar no longer increases monotonically with the magnetic inclination if the pulsar is surrounded by a boundary layer geometry. To prove this arguments, NSs with the same boundary layers and accretion disks as MAXI J1816--195 but different magnetic and observing inclinations are simulated. The X-ray pulse fractions of these NSs are shown in Figure~\ref{fig:pf_m_i}; the energy range is chosen as $1-10$ keV, in which most of AMXPs are observed and found. At many observing inclinations, the pulse fraction of NSs first increases and then decreases with magnetic inclination, suggesting that the possibility of a pulsar with large magnetic inclination but small pulse fraction cannot be excluded. Thus, another model different from the nearly aligned moving spot model of AMXPs is needed to interpret the rarity of AMXPs. Besides, the high pulse fractions at observed inclination $90^\circ$ in Figure~\ref{fig:pf_m_i} are because that the disk flux at viewing direction parallel to the disk is really small; as indicated by the corresponding pulse amplitudes shown in Figure~\ref{fig:amplitude_m_i}, the pulses at observing inclination $\sim 90^\circ$ are not strong.

\begin{figure}
\includegraphics[width=9cm]{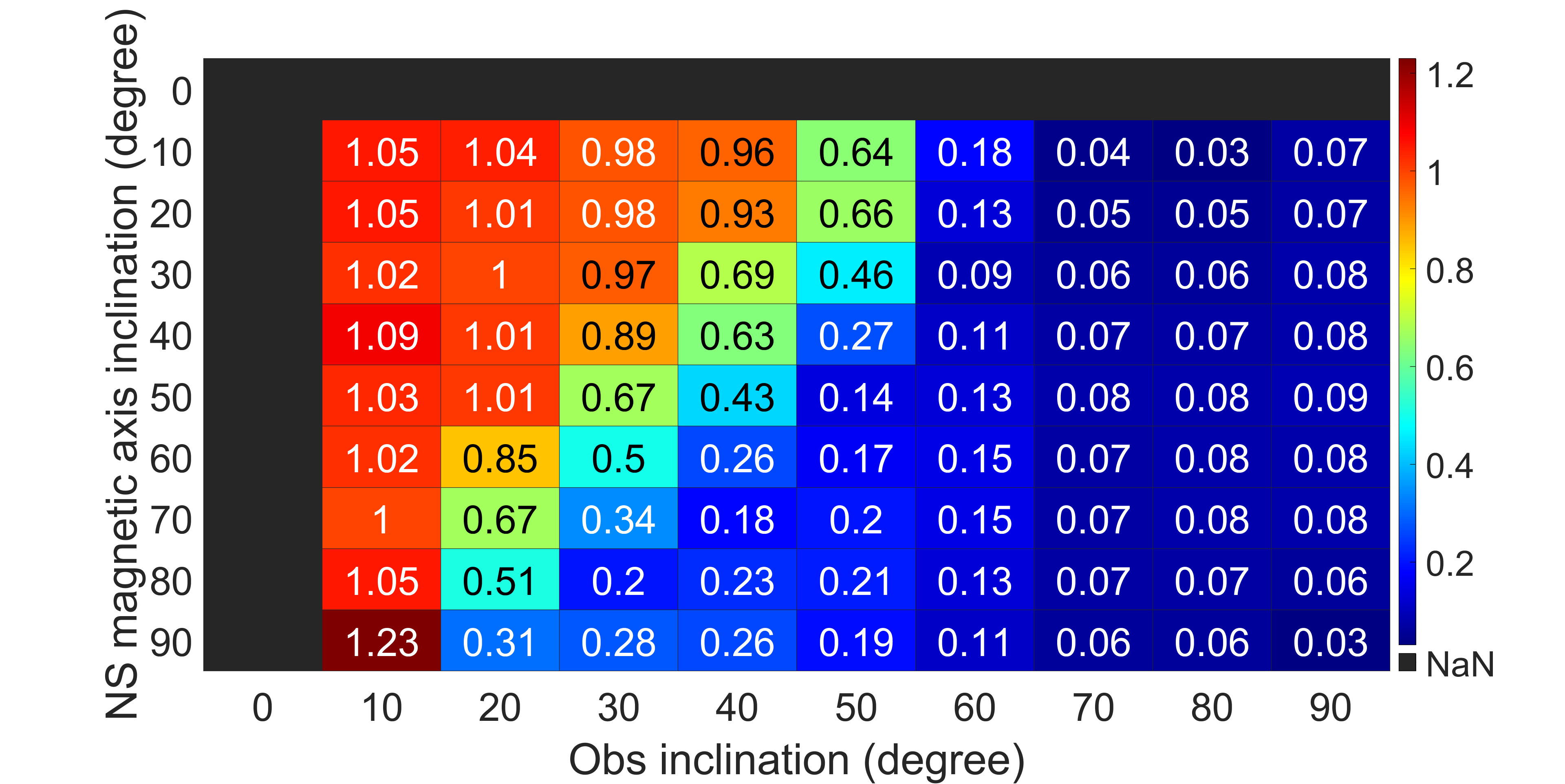}
\caption{Ratios between the simulated pulse amplitudes of MSPs with boundary layers surrounded and pulse amplitudes of isolated MSPs in energy range of $1-10$ keV, in cases of different observing inclinations and NS magnetic inclinations. Ratio less than 1 means that the pulsation of MSP is smeared by the boundary layer.}
\label{fig:ratio_m_i}
\end{figure}

It is indicated by Figure~\ref{fig:amplitude_m_i} that the pulses are strong when the X-ray beam from the NS magnetic pole sweeps over the edge of the boundary layer without being heavily obscured. To estimate the influence of the scattering in the boundary layer on the pulsation of NS radiation further, the ratios between the pulse amplitudes of NSs surrounded by the boundary layer geometry and the pulse amplitudes of isolated NSs are calculated and shown in Figure~\ref{fig:ratio_m_i}. It is shown that the intrinsic pulsation of a NS will not be smeared by the boundary layer only when both the observing and the magnetic inclinations of the NS are not large. The simulations in Figure~\ref{fig:ratio_m_i} are based on a boundary layer whose size is suitable for MAXI J1816-195; for other NS LMXBs, the optical depth and shape of the boundary layer may change, leading to different smearing level and angular range of smearing. However, it is evident that the pulse amplitude ratio shows a clear division between the smeared and not smeared regions, similar to Figure~\ref{fig:ratio_m_i}. Considering the distributions of the magnetic and observing inclinations of NSs, as a result, only a small part of NSs can be observed with pulsations when boundary layers exist. Therefore, if the boundary layer geometry is common in LMXBs, the rarity of AMXPs can be explained by the smearing of their boundary layers, at least to some extend.

It should be noted that our alternative explanation of the rarity of AMXPs with the boundary layer smearing model does not rule out the nearly aligned moving spot model for all AMXPs. The latter can still explain some of AMXPs whose pulse fractions are small and pulse profiles are sinusoidal. In addition, the soft states observed in the outbursts of some NS LMXBs are consistent with the nearly aligned moving spot model, but cannot be explained by the boundary layer smearing model. No significant non-thermal hard X-ray emission is detected in the soft state, suggesting that the boundary layer either does not exist or at least does not intercept the NS surface emission significantly. However, if the magnetic and spin axes of the NSs in these LMXBs are indeed aligned, these NSs cannot generate pulsed radiation because of their aligned magnetic axis, after the accretion stops or they become isolated NSs. Therefore these NS LMXBs with aligned magnetic and spin axes have no relation with MSPs. On the other hand, our boundary layer smearing model allows NS LMXBs to evolve into MSPs.

\subsection{Two predictions}

There are two predictions that can be used to test this boundary layer smearing model. One is that most AMXPs should have single-peaked pulse profile in low energy X-ray range, because in the boundary layer smearing model AMXPs are LMXBs with small observing and magnetic inclination angles, so that only radiation of the north magnetic pole can be seen directly, forming the single peak at low energy X-ray range. Among all known AMXPs, the vast majority of pulse profiles are single-peaked \citep{AMXP_review1, AMXP_review2}, suggesting that most AMXPs are consistent with the boundary layer smearing model; only three sources (Swift J1756.9--2508, see \citealt{1756Li2021}; SWIFT J1749.4–-2807, see \citealt{1749Altamirano2011}; XTE J1807--294, see \citealt{1807Kirsch2004}) show double-peaked shapes in low energy X-ray range. These three exceptions may be due to the insufficient optical depth of their boundary layers: these three sources may have large observing inclination angles, but their thin boundary layers fail to smear their radiations to make them be LMXBs without pulsations. Therefore the radiations of both the two magnetic poles are observed, forming the double-peaked pulse profiles. As for pulse profiles at high energy X-ray range, the formation of double-peaked shape needs a boundary layer with proper optical depth. Other than MAXI J1816--195, the X-ray pulse features of single-peaked at lower energy and double-peaked at high energy are also observed in IGR J18245--2452 \citep{18245DeFalco2017} and SAX J1808.4--3658 \citep{1808Ibragimov2009}. The lack of detection of double-peaked high energy X-ray pulsation in most AMXPs may be due to the inadequate sensitivity of most X-ray telescopes at high energies.

\begin{figure}
\includegraphics[width=9cm]{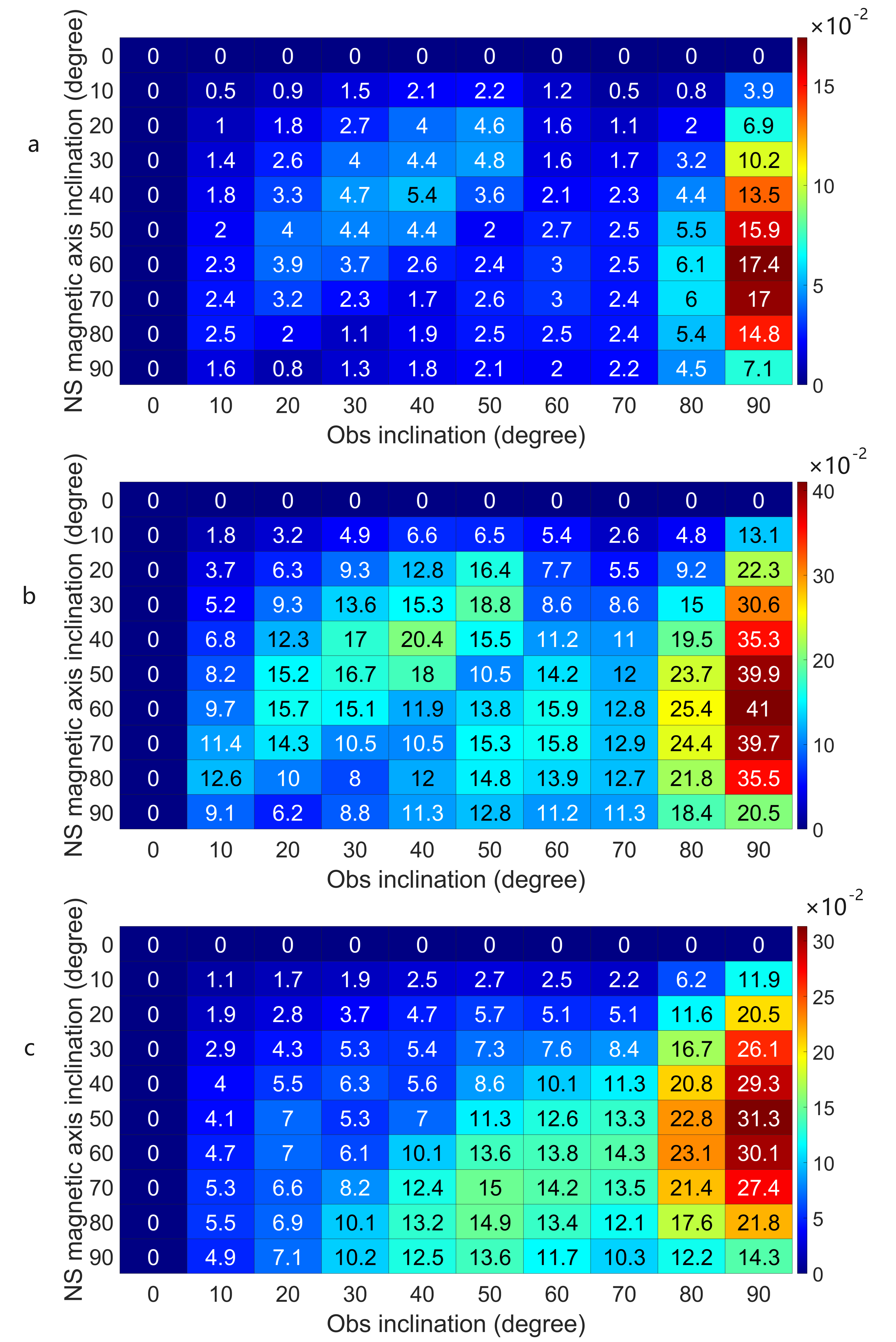}
\caption{The simulation results of pulse fraction in cases of different magnetic inclinations and observing inclinations, with energy range $1-10$ keV, $10-30$ keV and $30-100$ keV in panels a, b and c, respectively; the other settings are the same as Figure~\ref{fig:pf_m_i}.}
\label{fig:pf_m_i_3bands}
\end{figure}

\begin{figure}
\includegraphics[width=9cm]{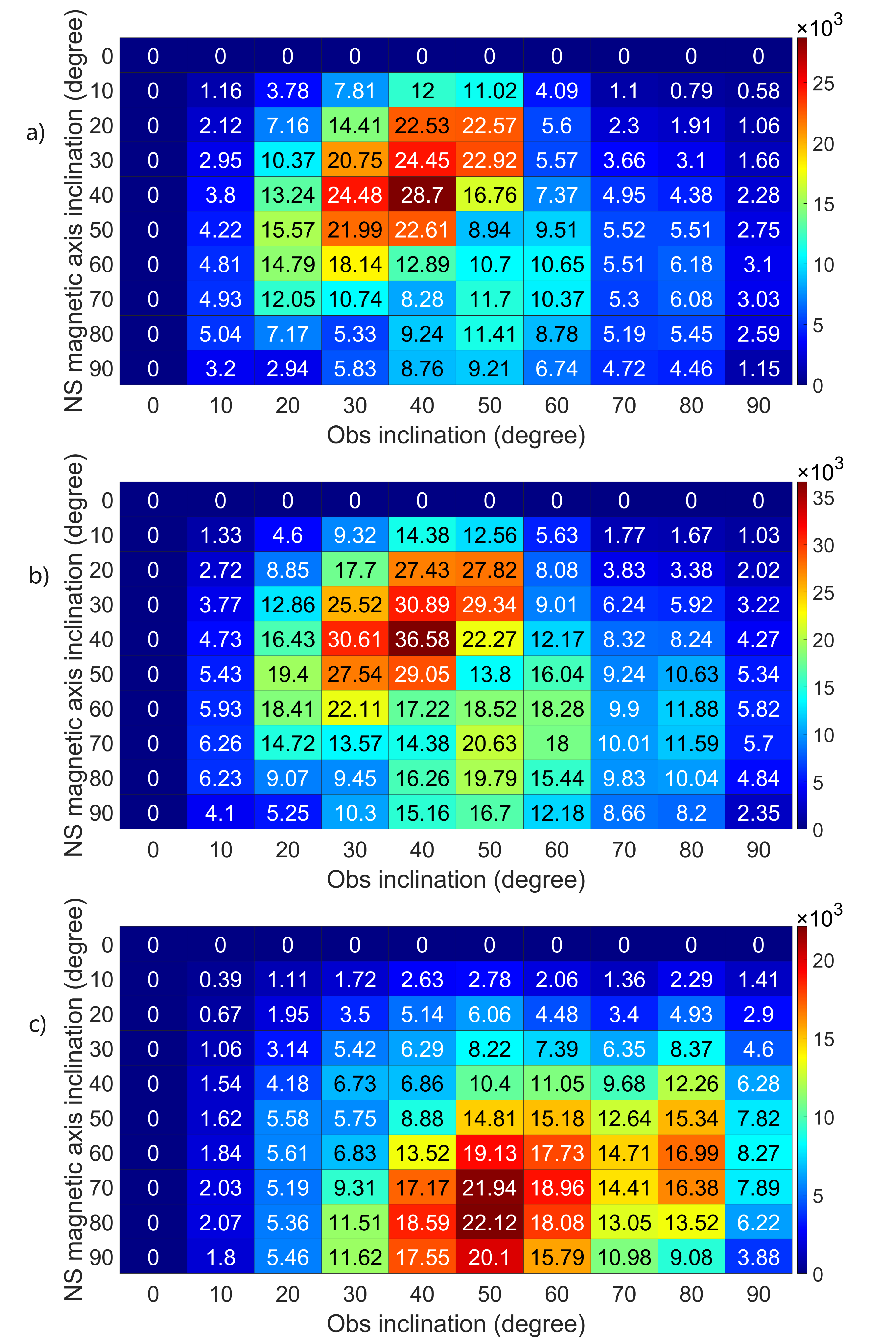}
\caption{The simulation results of pulse amplitude in cases of different magnetic inclinations and observing inclinations, in units of simulated photon number, with energy range $1-10$ keV, $10-30$ keV and $30-100$ keV in panels a, b and c, respectively; the other settings are the same as Figure~\ref{fig:amplitude_m_i}.}
\label{fig:amplitude_m_i_3bands}
\end{figure}

The other prediction of the boundary layer smearing model is that many LMXBs without obvious pulse at low energy X-ray range are expected to have observable pulsations at high energy X-ray range. From the simulation results of Figure~\ref{fig:pf_m_i_3bands} and Figure~\ref{fig:amplitude_m_i_3bands}, we can see that the pulse fractions and the pulse amplitude in higher energy ranges (panels b and c) are much larger than pulses in $1-10$ keV (panel a), and in $30-100$ keV the pulse amplitudes of the NSs obscured by their boundary layers are even larger than those of NSs without obscuration. There are two reasons that cause this huge change of pulsation at different energy ranges. On one hand, low energy X-ray radiation is mostly composed of the disk radiation, which has no pulsation and thus lowers the pulse fraction, but for high energy X-ray radiation the proportion of disk radiation becomes very small. On the other hand, some pulsed photons from the NS magnetic poles are scattered by the boundary layer and become pulsed high energy photons, so that the pulse fractions in high energy range are increased. Thus, it is likely for some NS LMXBs to have almost no pulse at low energy range but show pulsation at high energy range. We suggest to search for the expected pulse signal of LMXBs at high energy X-ray range to test the boundary layer smearing model. However, this pulse searching could be very hard because the high energy X-ray pulse flux of LMXBs are smaller than in low energy range, and there is no low energy X-ray pulse as the template for the folding of the high energy X-ray light curves; in this case, a new method of directly searching the high energy X-ray pulse is needed.

It seems that the scenario of an optically thick corona smearing the pulsation is in contraction to the low optical depth according to some spectral fitting \citep{challenge_corona_smearing}. However, the available shapes of the corona in XSPEC fitting, usually slab or spherical shape, sometimes are far from the actual coronae around NSs, so that the optical depth of the corona generated by XSPEC may not reflect the real optical depth. For instance, for the same spectrum the optical depth of the corona in MAXI J1816--195 given by XSPEC fitting is around 2.76 \citep{1816Li2023}; however, our MC simulation requires a boundary layer with an optical depth around 5 in radial direction. It is implied that the XSPEC-fitted optical depth may be underestimated if the corona is in the hollow cylinder shape rather than a slab or spherical shape assumed in XSPEC.

\subsection{Problems in the measurement of the mass and radius of accretion-powered MSPs}

\begin{figure}
\includegraphics[width=8.5cm]{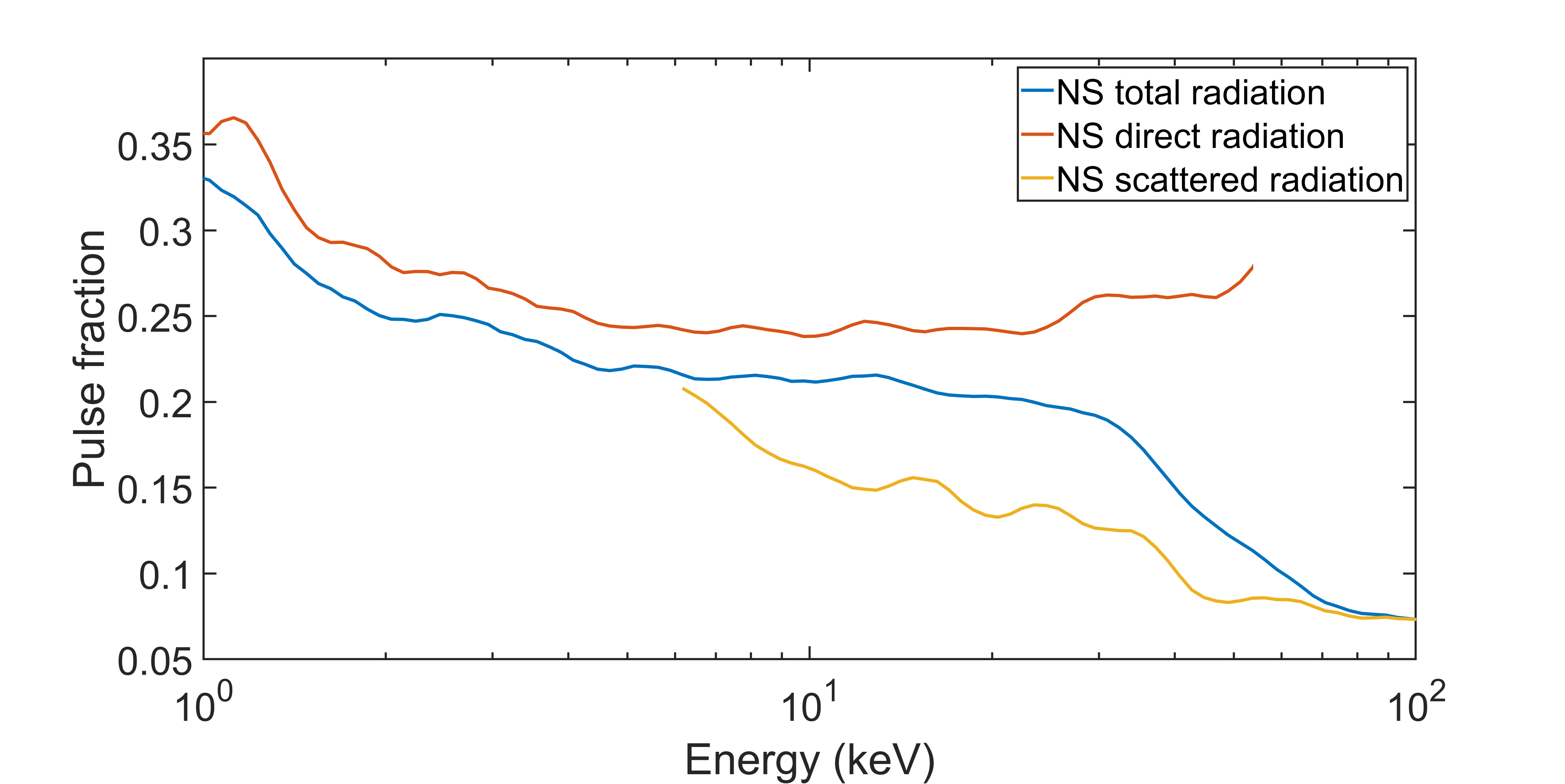}
\caption{The MAXI J1816--195 simulated PF of total radiation (blue line), radiation directly from the NS surface (red line) and scattered NS radiation (yellow line) as a function of energy. The PF of each component is calculated based on its own flux and only the PF with adequate statistics is plotted.}
\label{fig:pf_compare01}
\end{figure}

As mentioned in Section~\ref{sec:intro}, the mass $M$ and radius $R$ of a millisecond pulsar can be measured by pulse profile modelling. In this method, the oblate shape of the NS, the Doppler effect of the X-ray sources on the NS surface and the general relativistic effect on propagation of light are all taken into considerations, so that the accuracy of the parametric pulse profile modelling can reach $\sim0.1\%$ compared with numerical computation. The time cost of calculating pulse profile with this model on a single core of modern CPU can be much less than 1 s, making this method well suited for large-scale statistical sampling runs and Bayesian analyses \citep{ppmodeling_nicer2}. As a result, with $\geq10^7$ counts collected from a MSP, the pulse profile modelling method can measure $M$ and $R$ of the NS with uncertainties of a few percent \citep{ppmodeling_oblate, ppmodeling_review}.

\begin{figure}
\includegraphics[width=8.5cm]{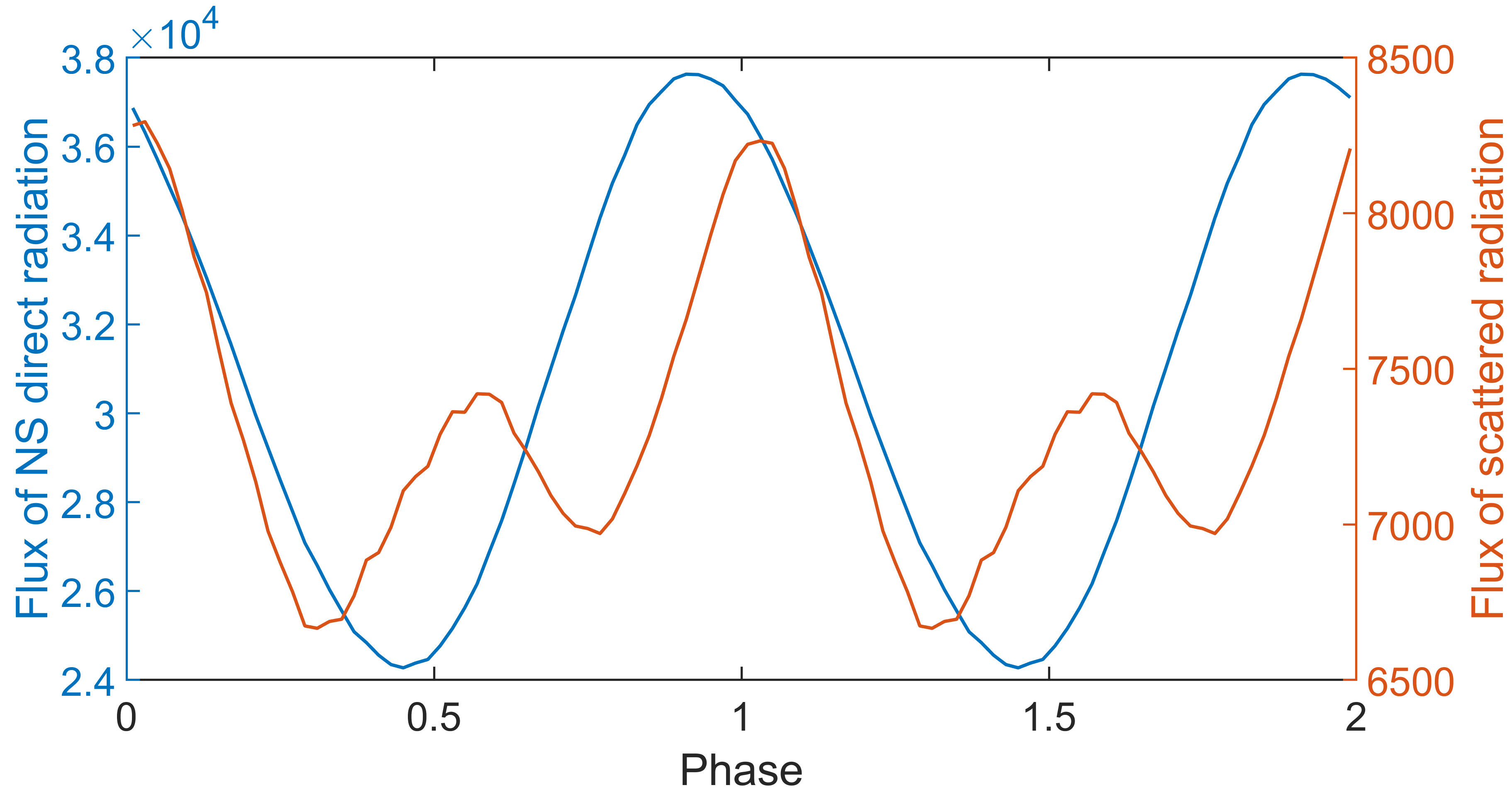}
\caption{The MAXI J1816--195 simulated pulse profile in $20-30$ keV of radiation directly from the NS surface (blue line) and scattered NS radiation (red line).}
\label{fig:pp_compare01}
\end{figure}

However, the pulse profile modeling method has difficulties on the $M$ and $R$ measurements of accretion-powered MSPs. This is partly because the accretion structures of NS, such as accretion disk and accretion flow corona, are neglected in the pulse profile modeling method \citep{ppmodeling_amxp}, but the radiation of these structure can seriously pollute the pulse profile. For instance, the radiation of MAXI J1816--195 mainly consists of disk radiation and scattered radiation, both of which are not directly from the NS surface, and this pollution makes huge influences on the observed pulse fraction and profile, as shown in Figure~\ref{fig:pf_compare01} and Figure~\ref{fig:pp_compare01}. Some works attempt to improve the pulse profile modeling and take the disk scattering into considerations \citep{ppmodeling_amxp_e1, ppmodeling_amxp_e2, ppmodeling_amxp}, leading to the increase of the parameter number of the pulse profile model. As a result, the parameters are degenerated with each other so that the radius $R$ of a NS can be constrained accurately only when the mass $M$ is fixed. On the other hand, the XSPEC model or empirical formula employed in these work for disk scattering are too simple to accurately describe the real scattering process happened around the NS.

As the arguments above, the pulse profile modeling method itself cannot deal with the $M$ and $R$ measurements of accretion-powered MSPs, and thus the influence of the scattering in coronae and the disk emission on the pulse profile should be estimated and eliminated at first. Our MC simulation is able to calculate these Compton scattering processes in details. In order to constrain the $M$ and $R$ of the accretion-powered MSPs, we will improve our MC simulation program, taking the oblate shape of the NS into consideration and calculating the general relativistic effects on the propagation of light. We hope that these future works can help to increase the sample size of the NS $M-R$ relations.

\begin{acknowledgments}

This work made use of the data and software from the Insight-HXMT mission, a project funded by China National Space Administration (CNSA) and the Chinese Academy of Sciences (CAS). This research has made use of data and software provided by data obtained from the High Energy Astrophysics Science Archive Research Center (HEASARC), provided by NASA’s Goddard Space Flight Center. This work is supported by the National Natural Science Foundation of China (Grant No. 12333007, 12027803, 12273030) and International Partnership Program of Chinese Academy of Sciences (Grant No.113111KYSB20190020).

\end{acknowledgments}

%

\vspace{5mm}


\bibliography{sample631}{}
\bibliographystyle{aasjournal}



\end{document}